\documentclass[10pt, conference]{IEEEtran}
\IEEEoverridecommandlockouts
\usepackage{cite}
\usepackage[table]{xcolor}
\usepackage{amsmath,amssymb,amsfonts}
\usepackage{algorithmic}
\usepackage{graphicx}
\usepackage{hyperref}
\usepackage{textcomp}
\usepackage{xcolor}
\def\BibTeX{{\rm B\kern-.05em{\sc i\kern-.025em b}\kern-.08em
    T\kern-.1667em\lower.7ex\hbox{E}\kern-.125emX}}
\usepackage{url}
\usepackage{makecell}
\usepackage{svg}
\usepackage{float}
\usepackage{fancybox}
\newenvironment{summarybox}%
{\begin{center}\vspace{0mm}\noindent\begin{Sbox}\begin{minipage}{0.97\columnwidth}}%
{\end{minipage}\end{Sbox}\fbox{\TheSbox}\end{center}\vspace{0mm}}
\begin{document}

\title{Understanding the Issue Types in Open Source Blockchain-based Software Projects with the Transformer-based BERTopic}

\author{
\IEEEauthorblockN{Md Nahidul Islam Opu}
\IEEEauthorblockA{
    \textit{SQM Research Lab} \\
    \textit{Computer Science} \\
    \textit{University of Manitoba}\\
    \textit{Winnipeg, Canada}
}
\and
\IEEEauthorblockN{Shahidul Islam}
\IEEEauthorblockA{
    \textit{SQM Research Lab} \\
    \textit{Computer Science} \\
    \textit{University of Manitoba}\\
    \textit{Winnipeg, Canada}
}
\and
\IEEEauthorblockN{Sara Rouhani}
\IEEEauthorblockA{
    \textit{TCDT Lab} \\
    \textit{Computer Science} \\
    \textit{University of Manitoba}\\
    \textit{Winnipeg, Canada}
}
\and
\IEEEauthorblockN{Shaiful Chowdhury}
\IEEEauthorblockA{
    \textit{SQM Research Lab} \\
    \textit{Computer Science} \\
    \textit{University of Manitoba}\\
    \textit{Winnipeg, Canada}
}
}

\maketitle

\begin{abstract}
Blockchain-based software systems are increasingly deployed across diverse domains, yet a systematic understanding of their development challenges remains limited. This paper presents a large-scale empirical study of 497,742 issues mined from 1,209 open-source blockchain projects hosted on GitHub. Employing BERTopic, a transformer-based topic modeling technique, we identify 49 distinct issue topics and organize them hierarchically into 11 major subcategories. Our analysis reveals that both general software development issues and blockchain-specific concerns are nearly equally represented, with \textit{Wallet Management} and \textit{UI Enhancement} emerging as the most prominent topics. We further examine the temporal evolution of issue categories and resolution times, finding that \textit{Wallet} issues not only dominate in frequency but also exhibit the longest resolution time. Conversely, \textit{Mechanisms} issues are resolved significantly faster. Issue frequency surged after 2016 with the rise of Ethereum and decentralized applications, but started declining after 2022. These findings enhance our understanding of blockchain software maintenance, informing the development of specialized tools and practices to improve robustness and maintainability.
\end{abstract}


\section{Introduction}

Blockchain technology has emerged as a transformative innovation with the potential to revolutionize a wide range of sectors \cite{bhutta_survey_2021} by offering enhanced transparency, security, and immutability. Beyond its foundational role in powering cryptocurrencies, blockchain is increasingly being adopted across diverse domains such as supply chain management \cite{balasubramanyam_adaptation_2020, kaur_adaptation_2022}, healthcare \cite{wrycza_analysing_2018}, and finance \cite{chang_how_2020}. The rapid growth of the cryptocurrency market, which has surpassed a trillion-dollar valuation \cite{zhou_cryptocurrency_2024}, has further accelerated interest and investment in blockchain-based solutions. This widespread adoption has led to a surge in the development of blockchain application projects, creating a vibrant and rapidly evolving ecosystem of blockchain software development.
Developing high-quality and maintainable applications requires a clear understanding of the nature and characteristics of software issues \cite{zou_empirical_2015, thung_empirical_2012} encountered during development, testing, and maintenance. These characteristics include the distribution of issues across different components \cite{uddin_empirical_2021}, their impact on the system, resolution times, and the correlation between various issue types. Gaining insights into these aspects can guide the prioritization of development and testing \cite{lal_blockchain_2021} efforts. For instance, if a significant number of issues arise from cryptographic security flaws, it would indicate the need for more specialized tools, deeper domain expertise, and a stronger emphasis on security-focused development practices. Such understanding enables more informed decision-making and contributes to the creation of more robust, secure, and maintainable applications.

Recent research has focused on identifying issues and bugs in blockchain-based systems, aiming to uncover the underlying characteristics and trends that developers encounter. However, these studies have either focused narrowly on bugs within a limited number of projects \cite{wan_bug_2017} or examined other types of development activities, such as commits and pull requests \cite{das_empirical_2022}. To the best of our knowledge, no prior research has undertaken a systematic, large-scale analysis of issue reports from blockchain-based applications on GitHub to comprehensively characterize the development challenges in this domain.

To address this gap, we collected 497,742 issues from a curated set of 1,209 open-source blockchain-based projects on GitHub and applied topic modeling techniques to uncover the key development challenges faced by practitioners in this domain. Through this large-scale empirical analysis, we aim to answer the following three research questions:

\textbf{RQ1: What are the key issue categories in blockchain projects as identified through topic analysis?}

By leveraging advanced topic modeling techniques using BERTopic on GitHub issues, followed by systematic manual annotation, we identified and categorized 49 distinct topics across the analyzed projects. Among these, the most dominant were \textit{Wallet Management and Connectivity} (21.3\%), \textit{UI Enhancement} (10.6\%), and \textit{Logging and Error Handling} (9.2\%).

\textbf{RQ2: How do resolution times vary across different subcategories of issues in blockchain projects?}

The resolution time analysis shows that \textit{Wallet} issues take notably longer to resolve, whereas \textit{Mechanisms} issues, such as those involving consensus protocols, the Network Nervous System (NNS), and neuron staking, are resolved more quickly.


\textbf{RQ3: How have different types of issues evolved over time in terms of frequency within blockchain projects?}

We examined the temporal trends in issue creation across key subcategories and identified a significant surge in reported issues beginning around 2016. This surge was particularly pronounced in categories such as \textit{Development \& Maintenance}, \textit{User Experience \& Interaction}, and \textit{Wallet}. 

To facilitate replication and extension, we publicly share our data and code \footnote{\scriptsize\url{https://github.com/SQMLab/StudyOnBlockchainApps}}.

\section{Background \& Related Works}
\label{sec:relatedWorks}

Numerous studies have explored issues, particularly from GitHub, to understand developer concerns, issue trends, and resolution patterns. Tamanna et al. \cite{tamanna_characterizing_2023} conducted a large-scale empirical study of 118K GitHub issues from 34 runtime system repositories to categorize issue types and analyze issue management practices. Beyer and Pinzger \cite{beyer_manual_2014} conducted a manual classification of Android development issues on Stack Overflow, uncovering recurring challenges faced by mobile developers. In addition to manual categorization, various automated approaches have been used to analyze issue-fixing times, particularly for bugs. For example, Lamkanfi et al. \cite{lamkanfi_filtering_2012} observed skewed bug-fix time distributions in Eclipse and Mozilla, while Panjer \cite{panjer_predicting_2007} applied logistic regression to predict bug resolution times using Eclipse Bugzilla data. Giger et al. \cite{giger_predicting_2010} advanced this line of work using decision-tree-based models, achieving improved precision and recall. More recently, Ardimento et al. \cite{ardimento_using_2020} applied BERT-based models to predict bug-fixing durations, reporting accuracy up to 91\%.

Topic modeling has been extensively applied to mine latent themes in developer communications, including GitHub, Stack Overflow, and issue trackers. Thung et al. \cite{thung_empirical_2012} and Uddin et al. \cite{uddin_empirical_2021} used Latent Dirichlet Allocation (LDA) to identify 40 prevalent topics in bug reports and developer Q\&A, respectively. Bangash et al. \cite{bangash_what_2019} explored how developer discussions evolve temporally in the machine learning domain on Stack Overflow, revealing trends and tagging challenges by identifying 50 topics. More recently, BERTopic has emerged as a state-of-the-art topic modeling method that combines transformer-based embeddings with clustering techniques to extract coherent and semantically meaningful topics. Grootendorst \cite{grootendorst_bertopic_2022} demonstrated that BERTopic outperforms classical models in coherence and human evaluation. Egger et al. \cite{egger_topic_2022} found BERTopic particularly effective for modeling short and noisy texts, such as Twitter data, demonstrating its versatility.

Several recent studies have applied topic modeling with GitHub issue analysis to study blockchain-related software development. Vaccargiu et al. \cite{vaccargiu_sustainability_2024} utilized BERTopic on GitHub issues and comments to uncover sustainability-related themes in Ethereum developer discussions. In a related study, Vaccargiu and Tonelli \cite{vaccargiu_blockchain_2024} applied BERTopic to blockchain projects in the environmental sector, identifying core concerns such as carbon credit tracking and resource optimization. Wan et al. \cite{wan_bug_2017} manually classified 946 unique bugs across eight blockchain systems into ten categories using card sorting and investigated bug-fixing dynamics in blockchain systems. Furthermore, Wan et al. \cite{wan_what_2021} analyzed blockchain-related discussions from Stack Overflow. Using LDA, they extracted 45 topics distributed across six architectural layers. Although this research shares objectives with ours, it diverges in its reliance on Q\&A platforms as the data source. Repository mining on GitHub, by contrast, offers a more direct lens into the processes of software development and evolution, as issue reports typically capture long-term maintenance challenges~\cite{kalliamvakou_promises_2014}, whereas Stack Overflow posts predominantly reflect short-term problem solving and knowledge sharing~\cite{treude_how_2011}. This difference between these two data sources motivated us to choose GitHub in our study.

Despite substantial progress in issue analysis, topic modeling, and blockchain software research, advanced topic modeling has not yet been systematically applied to GitHub issues in the blockchain domain. Prior studies either address general software contexts or focus on specific concerns, leaving a gap in providing a comprehensive, data-driven view of developer-reported challenges. This study seeks to address that gap.

\section{Methodology}
\label{sec:methodology}
The methodology is organized into two principal phases: (i) data collection and preprocessing of GitHub repositories and issues, (ii) topic modeling using BERTopic to uncover recurring topics. Each phase is described in detail below.

\subsection{Data Collection}

\begin{figure}[ht]
    \centering
    \includegraphics[width=0.93\linewidth]{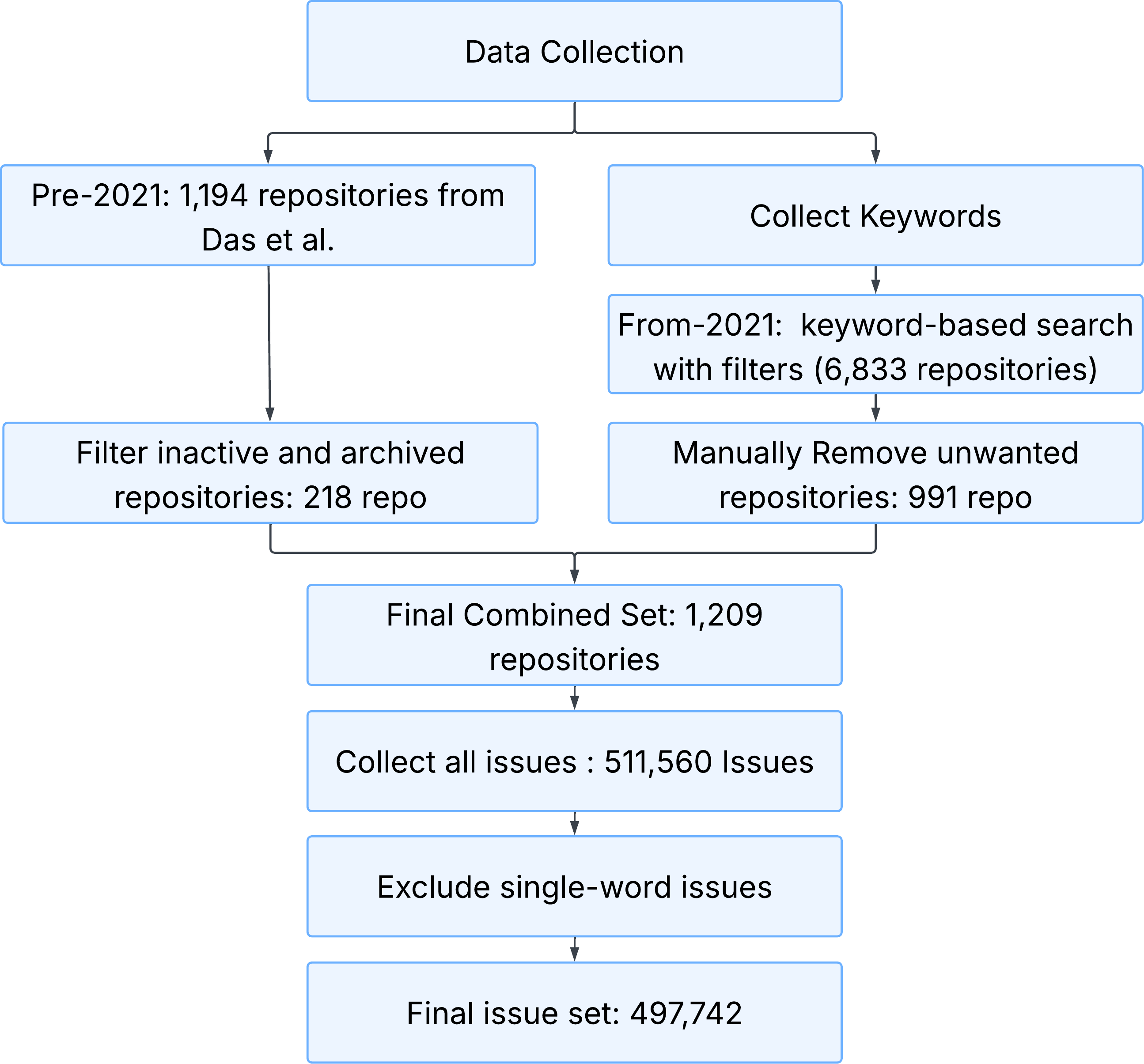}
    \caption{Steps in data collection process.}
    \label{fig:data_collection}
\end{figure}

GitHub is a key platform for software artifacts, hosting millions of repositories with rich metadata (e.g., pull requests, issues, commits), offering valuable empirical insight into large-scale software development practices~\cite{kalliamvakou_promises_2014}. We therefore selected GitHub as our primary source for blockchain-related repositories. Figure~\ref{fig:data_collection} outlines the data collection pipeline.

\subsubsection{Project Selection}



Identifying blockchain-related open-source repositories on GitHub is challenging due to inconsistent labeling and keyword usage, leading to search inaccuracies~\cite{das_empirical_2022}. To address this, we adopted a semi-automated, two-phase selection strategy.

First, for repositories created before January 2021, we used the curated dataset from Das et al.\cite{das_empirical_2022}, based on a validated GHTorrent subset~\cite{gousios_ghtorent_2013}. This dataset originally included 1,194 repositories filtered by cryptocurrency relevance, platform affiliation, and popularity. However, as their dataset represents a snapshot up to January 2021, many of these repositories are found to be inactive or archived. Applying additional filtering to remove such projects, we retained 218 active repositories.

Second, since GHTorrent was discontinued after 2021, to collect the new repositories, we conducted a systematic query using the \texttt{PyGithub}\footnote{\scriptsize\url{https://github.com/PyGithub/PyGithub}} Python library that uses GitHub REST API\footnote{\scriptsize\url{https://docs.github.com/en/rest}}. The keyword list used in the search was adapted from Das et al.'s methodology~\cite{das_empirical_2022}: 

\begin{itemize}
    \item[(i)] names of major platforms and cryptocurrencies (e.g., Bitcoin, Ethereum), extracted from CoinMarketCap\footnote{\scriptsize\url{https://coinmarketcap.com}};
    \item[(ii)] names of prominent permissioned blockchain frameworks (e.g., Hyperledger Fabric, Corda), which lack cryptocurrency yet represent significant enterprise use cases; and
    \item[(iii)] blockchain-related terms (e.g., smart contracts, dApps, Solidity) to capture relevant tools and applications.
\end{itemize}

To further refine the keyword set, popular decentralized application (dApp) names were incorporated based on rankings and traffic data from The Dapp List\footnote{\scriptsize\url{https://thedapplist.com}}, a curated directory highlighting trending dApps across multiple blockchains, and DappRadar\footnote{\scriptsize\url{https://dappradar.com}}, a widely-used analytics platform that tracks dApp usage statistics across blockchain networks. Including these sources helped ensure coverage of emerging and actively maintained projects that might otherwise be overlooked through generic keyword searches alone.

Additional automated filtering was applied to reduce noise: repositories were excluded if empty, archived, inactive in the past year, or had fewer than five stars or forks. These criteria, based on empirical guidelines~\cite{das_empirical_2022, kalliamvakou_promises_2014}, aimed to balance dataset inclusiveness and quality. Inactivity served as a proxy for abandonment, while stars and forks indicated interest and collaboration. After filtering, 6,833 repositories remained.

\subsubsection{Project Filtering}
Mining GitHub repositories presents several well-documented challenges. As highlighted by Kalliamvakou et al.~\cite{kalliamvakou_promises_2014}, many do not represent actual software projects but instead host tutorials, assignments, or code snippets. Consequently, we undertook a comprehensive manual verification process to refine the initially collected 6,833 repositories. Each repository was evaluated individually, and those meeting any of the following conditions were excluded: (i) not related to blockchain software development; (ii) a demo application; (iii) used primarily for educational purposes (e.g., tutorials, course assignments, or coding exercises); (iv) serving as a static data or resource storage.

In summary, we excluded repositories not representing genuine blockchain software systems, resulting in 991 high-confidence repositories after 80+ hours of manual validation by the first two authors. Combined with 218 repositories from Das et al.~\cite{das_empirical_2022}, this yielded a final dataset of 1,209 repositories.

\subsubsection{Issue Collection}
Following the identification of the target repositories, we collected all associated issues using the GitHub REST API via the \texttt{PyGithub} library. Notably, we discovered that 185 repositories contained no reported issues. Overall, we collected a total of 511,560 issues, encompassing both open and closed issues, without filtering by type.
 
Upon conducting a manual inspection of the issue corpus, we observed that a substantial number of issues consisted of a single word, such as \textit{Develop}, \textit{Dev}, \textit{Development}, \textit{Master}, \textit{Test}, \textit{Staging}, \textit{Backup}, \textit{Release}, etc., or a version number. These issues do not provide any information on the topic of the issue, but rather create ambiguity and lack analytical value. Consequently, we excluded all single-word issues from the dataset. After this step, we retained 497,742 issues, which served as the basis for the subsequent stages of our study.

\subsection{Topic Modeling with BERTopic}

To identify and analyze topics within issues of blockchain software repositories, we employed BERTopic, a modern topic modeling framework that combines transformer-based embeddings, dimensionality reduction, and hierarchical clustering \cite{grootendorst_bertopic_2022}. Given its superior performance over traditional methods~\cite{egger_topic_2022, gan_experimental_2024, medvecki_multilingual_2024, el-gayar_comparative_2024, kaur_moving_2024}, BERTopic was selected for this task.

\begin{figure} [!h]
    \centering
    \includegraphics[width=0.65\linewidth]{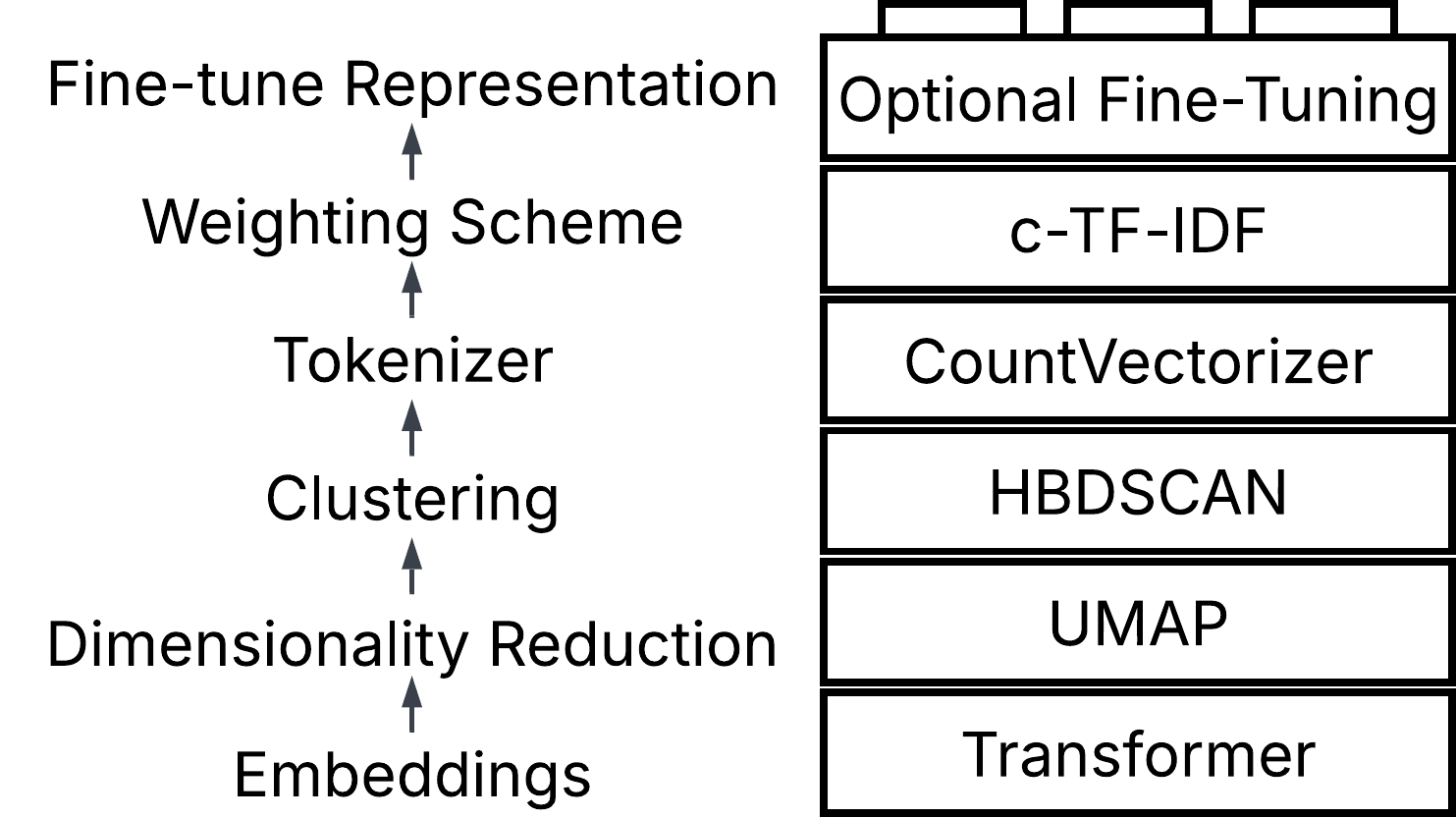}
    \caption{BERTopic~\cite{grootendorst_bertopic_2022} can be viewed as a sequence of 6 steps to create its topic representations. The first three steps are for creating the topics, and the last three steps are for representing and interpreting the topics.}
    \label{fig:enter-label}
\end{figure}

BERTopic offers extensive customization, requiring iterative tuning. Starting with default settings, we observed over a thousand imbalanced topics—some overly broad, others too narrow. This motivated a trial-and-error approach to refine parameters and optimize topic coherence.

BERTopic begins by converting texts into embeddings using transformer models. Unlike traditional topic modeling, which requires extensive data preprocessing~\cite{makhshari_iot_2021, bangash_what_2019}, it requires minimal preprocessing. As noted in the official BERTopic documentation\footnote{\scriptsize\url{https://maartengr.github.io/BERTopic/faq.html?\#should-i-preprocess-the-data}}, traditional preprocessing is unnecessary when using embeddings, as every part of a document contributes to understanding its overall topic. But, it is recommended to clean the text of excessive noise, such as HTML tags, as these do not contain contextual information. Following this, we removed only markup tags, emojis, and URLs from the data.

For numerical representation, we selected the \textit{all-mpnet-base-v2} model from the SentenceTransformers library. This model balances state-of-the-art embedding quality with computational efficiency and has been shown to perform robustly across a variety of Natural Language Processing tasks~\cite{wang_identifying_2024}. 

We initially used concatenated issue titles and bodies for topic modeling, but this led to noisy, semantically diffused clusters with thousands of topics. Analysis revealed that many bodies were auto-generated by GitHub Actions or CI/CD tools or template-based, offering little useful content—an issue also noted in prior studies~\cite{li_follow_2023, sulun_empirical_2024}. To improve topic quality and interpretability, we limited modeling to issue titles, which are typically concise and semantically rich.

After generating an embedding, the next step is to cluster the embeddings. To facilitate faster clustering, we applied Uniform Manifold Approximation and Projection (UMAP) for dimensionality reduction. This choice is justified by UMAP’s ability to preserve global semantic relationships while enhancing local cluster separability~\cite{mcinnes_umap_2020}. We retained 100 dimensions and set the neighborhood size to 25, balancing computational efficiency with contextual richness in high-volume data. These parameters were chosen through trial and error, as the combination of high data volume and numerous parameters made a grid-search type approach challenging.

Clustering was performed using Hierarchical Density-Based Spatial Clustering of Applications with Noise (HDBSCAN), a density-based algorithm well-suited for discovering arbitrarily shaped clusters and ignoring outliers~\cite{campello_density-based_2013, asyaky_improving_2021}. A minimum cluster size of 1,000 was set to capture dominant issue themes and reduce noise, improving topic label clarity. Initial experiments showed that smaller values produced thousands of topics, hindering manual analysis. Although this threshold resulted in 268,755 outliers, we ignored the outliers, prioritizing prominent and recurring topics, excluding infrequent ones.

After topic generation, BERTopic refines topic representation through three steps: c-TF-IDF with CountVectorizer to reduce frequent terms, and a hybrid approach combining KeyBERT-inspired extraction with Maximal Marginal Relevance (MMR) to enhance relevance and diversity. This process helps in providing meaningful names to each topic.

Finally, the selected BERTopic model configuration, with automatic topic count selection and probability estimation, yielded 49 distinct topics—a similar number to the 50 and 40 topics identified by Bangash et al. \cite{bangash_what_2019} and Uddin et al. \cite{uddin_empirical_2021}, respectively, as noted in Section~\ref{sec:relatedWorks}.


\section{Approach, Analysis \& Results}
\label{sec:results}



In this section, we answer three research questions (RQ) based on the 49 topics generated by BERTopic. 
\subsection*{\textbf{RQ1:} What are the key issue categories in blockchain projects as identified through topic analysis?}

\begin{figure*}[!ht]
    \centering
    \includegraphics[height=8.5cm, width=1\linewidth]{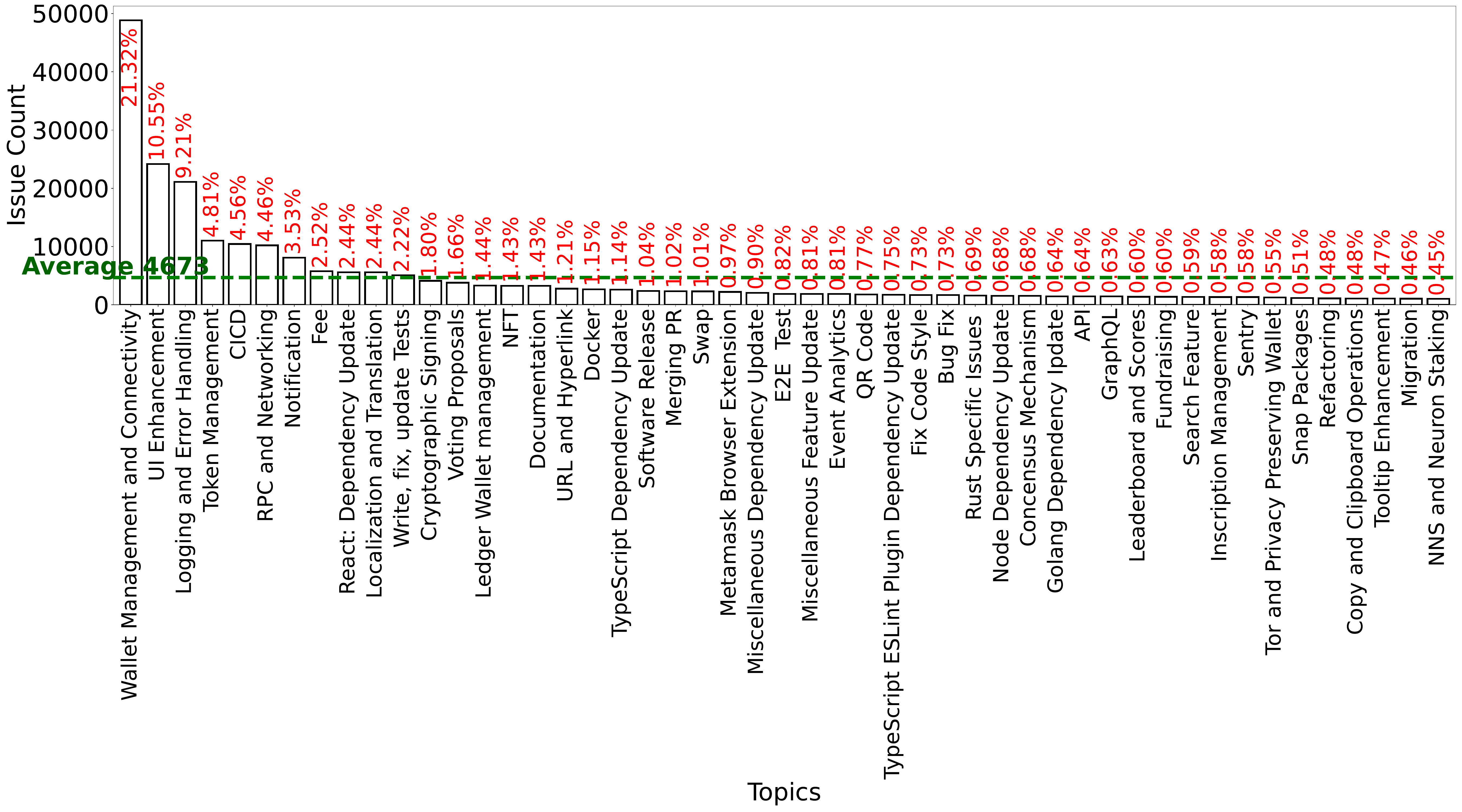}
    \caption{49 topics are generated using BERTopic.}
    \label{fig:49_topics}
\end{figure*}

After generating the topics, the next critical step involved assigning semantically meaningful and representative labels to each topic to facilitate interpretation. This labeling was performed collaboratively in real-time by the first and second authors, both researchers with 3 and 8 years of expertise in software engineering and testing, through an iterative process of discussion and refinement. For each topic, 30 highly representative issues were examined, ranked based on their probability values, to determine the most suitable label. The process continued until both researchers reached a consensus.
Figure~\ref{fig:49_topics} illustrates the distribution of issue percentages across all 49 identified topics. The top five most prominent topics identified across all issues were: \textit{Wallet Management and Connectivity} (21.3\%), \textit{UI Enhancement} (10.6\%), \textit{Logging and Error handling} (9.2\%), \textit{CICD} (4.6\%), and \textit{RPC and Networking} (4.5\%). The high volume of UI-related issues likely reflects the complexity of blockchain technology and unintuitive interfaces. Even popular wallets pose challenges: new users struggle with complex interactions~\cite{gandhi_usability_2022}, while experienced users face unfamiliar metaphors and transaction flows~\cite{saldivar_blockchain_2023}. Our analysis supports this, with frequent terms like icon, modal, and mobile in issues.

\begin{figure*}[p]
    \centering
    \includegraphics[height=0.95\textheight]{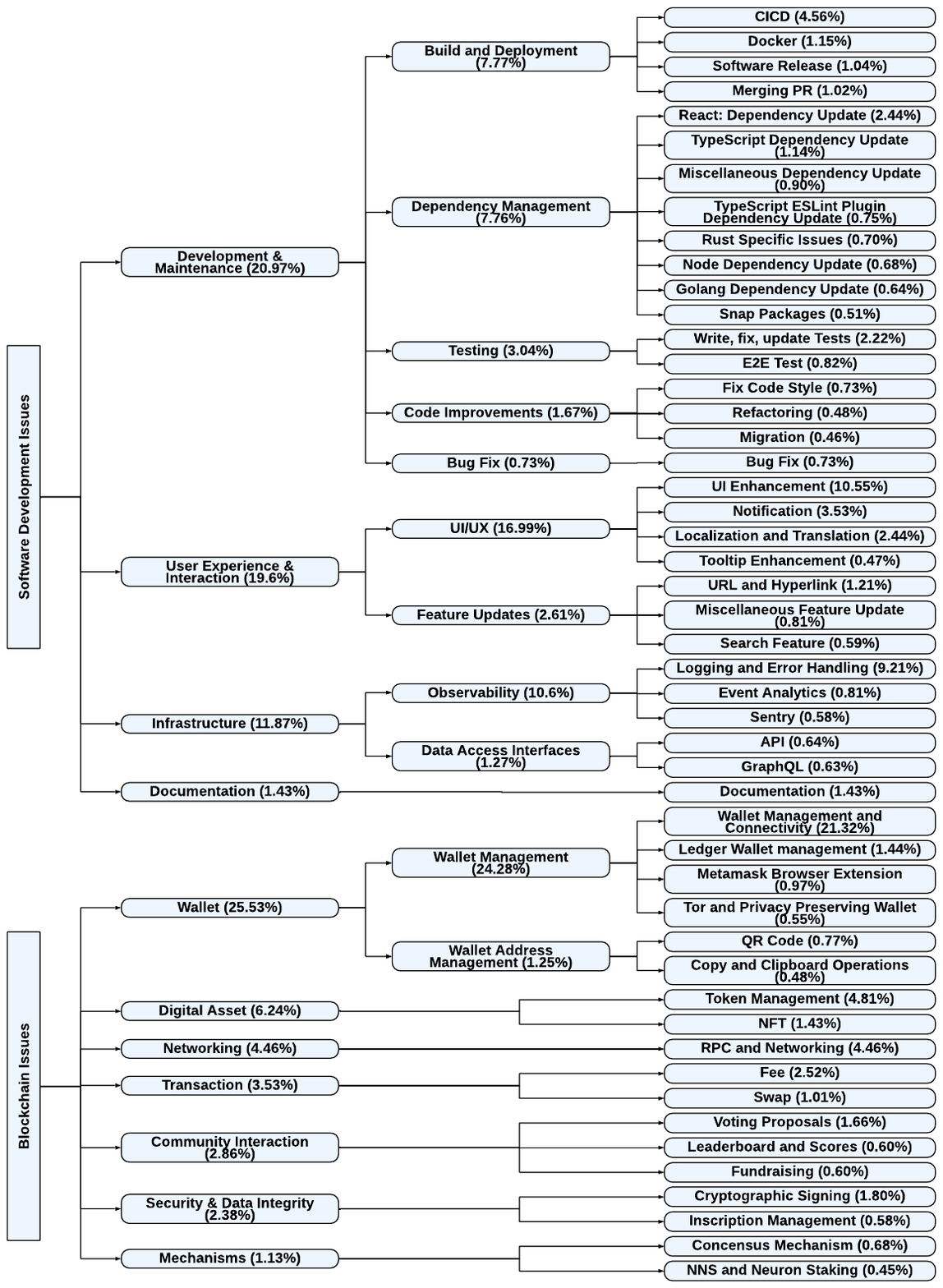}
    \caption{Hierarchy of topics with 11 major subcategories.}
    \label{fig:hierarchy}
\end{figure*}

After labeling, the topics were recursively aggregated into higher-level subcategories by merging semantically and conceptually related topics, yielding 11 subcategories at the intermediate level and two categories at the root of the hierarchical structure. This process was guided by domain knowledge and aligned with practices reported in prior research \cite{bagherzadeh_going_2019, yang_what_2016}. For example, topics such as \textit{Write, Fix, Update Tests} and \textit{E2E Test} were placed under the broader subcategory of \textit{Testing}. This categorization resulted in two overarching categories: \textit{Software Development Issues} (53.87\%) comprising 4 primary subcategories, and \textit{Blockchain-Specific Issues} (46.13\%) consisting of 7 primary subcategories. The hierarchical structure of these topics is illustrated in Figure \ref{fig:hierarchy}.


From the constructed hierarchy, we selected 11 major subcategories to guide our analysis for the subsequent research questions. We provide a high-level summary of the 11 major subcategories, each accompanied by a brief description.


\textbf{Development \& Maintenance.} This is the largest subcategory under \textit{Software Development Issues} category, comprising 18 topics and accounting for 20.97\% of all issues. These issues reflect the day-to-day challenges developers face in keeping software systems reliable and maintainable—such as resolving build failures, managing conflicting dependencies, addressing flaky tests, refining code quality, and fixing functional bugs.

\textbf{User Experience \& Interaction.} This is the second largest subcategory under \textit{Software Development Issues} category, comprising 19.6\% of all issues. It includes seven topics that address design inconsistencies, language localization, translation issues, and notification features aimed at improving overall user interaction.

\textbf{Infrastructure.} One major topic of this subcategory, \textit{Observability} (24,278 issues; 10.6\%), includes concerns related to \textit{Logging and Error Handling}, \textit{Event Analytics}, and tools like \textit{Sentry}, focusing on actionable logs, error tracking, diagnostics, and analytics integration. Meanwhile, the \textit{Data Access Interfaces} subcategory addresses query efficiency, endpoint design, and schema evolution—highlighting the foundation for performance monitoring, debugging, and system integration.

\textbf{Documentation.}  Comprising a single topic, this subcategory accounts for 3,267 (1.43\%) issues of the total. These issues often address missing or outdated instructions, unclear usage guides, API documentation gaps, and requests for improved onboarding materials. 

\textbf{Wallet.} Encompassing six topics and representing 25.53\% of all issues, this subcategory is divided into two low-level subcategories. \textit{Wallet Management} (24.28\%) covers discussions related to managing wallet ledgers, \textit{MetaMask Browser Extensions}, \textit{Tor, and Privacy Preserving Wallets}. Meanwhile, \textit{Wallet Address Management} (1.25\%) focuses on issues involving QR code, address copying, and clipboard operations.

\textbf{Digital Asset.} This subcategory accounts for 6.24\% of issues, primarily covering \textit{Token Management} (4.81\%) and \textit{Non-Fungible Tokens (NFTs)} (1.43\%). Token Management issues typically involve minting, transferring, balance synchronization, and integration with multi-chain or custom tokens. NFT-related discussions focus on metadata rendering, display inconsistencies, and listing functionality

\textbf{Networking.} Most issues in this subcategory relate to RPC failures, latency, and sync errors—often involving timeouts, broken connections, and unreliable data retrieval—affecting responsiveness and reliability in network-dependent tasks.

\textbf{Transaction.} The \textit{Transaction} subcategory represents 3.53\% of all issues (8,084 in total), encompassing concerns around transaction fees (2.52\%) and token swap functionality (1.01\%). Common issues include miscalculated or unexpectedly high gas fees, failed or delayed transactions, and poor user feedback during execution.

\textbf{Community Interaction.} This subcategory encompasses three distinct topics, accounting for 2.86\% of all issues. These topics include \textit{Leaderboard and Scores} (0.60\%), which focuses on reputation tracking, \textit{Fundraising} (0.60\%), which deals with community-driven financial support, and \textit{Voting Proposals} (1.66\%), which highlights governance mechanisms.

\textbf{Security \& Data Integrity.} This subcategory accounts for 2.38\% of all issues (5,463 total), focusing on core cryptographic operations in blockchain systems. It includes \textit{Cryptographic Signing} (1.8\%), covering digital signatures and authentication workflows, and \textit{Inscription Management} (0.58\%), addressing secure on-chain data encoding and handling.

\textbf{Mechanisms.} It encompasses two topics: \textit{Consensus Mechanism} (0.68\%), which covers the rules and algorithms by which network nodes agree on the blockchain state, and \textit{Network Nervous System (NNS) and Neuron Staking} (0.45\%), which addresses governance and staking operations within the network’s neuron-based voting system.

These results indicate that wallet-related functionality, which is unique to blockchain ecosystems, represents the largest proportion of issues, and the second largest subcategory consists of software development and maintenance issues. 

\begin{figure*}[ht]
    \centering
    \includegraphics[width=0.85\textwidth]{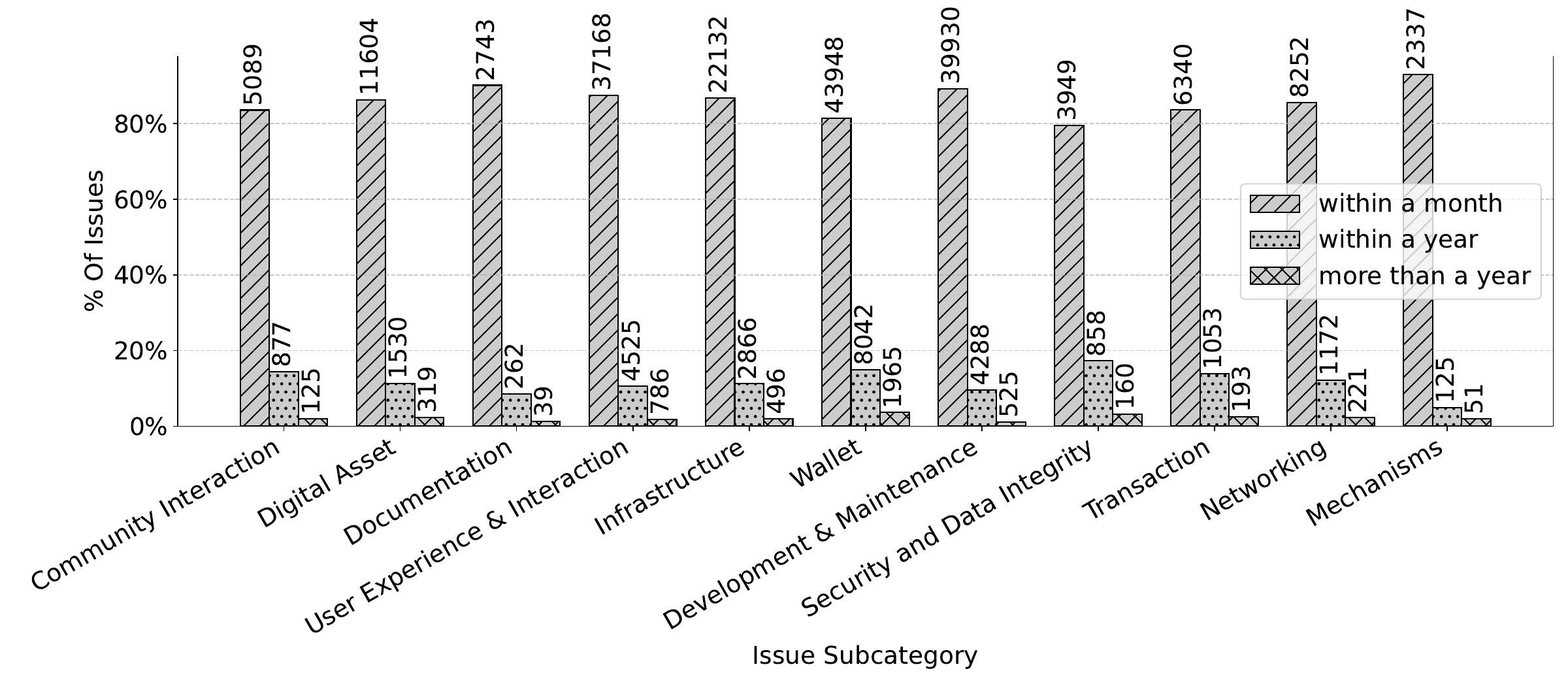}
    \caption{Percentage of issues resolved within different timeframes across subcategories, showing consistent resolution patterns.}
    \label{fig:resolution_distribution}
\end{figure*}

\begin{summarybox}
\textit{Wallet Management and Connectivity}, \textit{UI Enhancement}, and \textit{Logging and Error Handling} are the most prevalent topics in blockchain projects. In hierarchical terms, \textit{Development \& Maintenance} and \textit{Wallet} are the dominant subcategories, with blockchain-specific and general development issues appearing in nearly equal measure.
\end{summarybox}


\subsection*{\textbf{RQ2:} How do resolution times vary across different subcategories of issues in blockchain projects?}
To examine how resolution times vary across subcategories, we computed the minimum, maximum, mean, median, and $10^{th}$ and $90^{th}$ percentile durations for each subcategory (Table~\ref{tab:resolution_times}). These statistics reveal notable differences in resolution patterns across issue types. For example, some issues are closed within minutes of creation. Instances like \href{https://github.com/MystenLabs/sui/issues/1180}{\textit{MystenLabs/sui/issues/1180}\footnote{\scriptsize\url{https://github.com/MystenLabs/sui/issues/1180}}} and \href{https://github.com/Vagabonds-Labs/cofiblocks/issues/100}{\textit{Vagabonds-Labs/cofiblocks/issues/100}} demonstrate this behavior. Prior studies have reported such cases where developers write a patch, commit it to the version control system, and then create and immediately close the corresponding issue \cite{wan_bug_2017, thung_empirical_2012, lamkanfi_filtering_2012}.

To determine whether the observed differences in resolution times across subcategories are statistically significant, we applied the non-parametric Kruskal-Wallis H-test, suitable for non-normally distributed data. The test yielded a highly significant result: \textit{\(H = 3640.67\), \(p = 0.0\)}, confirming that resolution times differ significantly across subcategories. 

\begin{table}[h]
\rowcolors{2}{white}{gray!15}
\centering
\caption{Resolution time (in days) of closed issues by subcategory. Reported values include minimum, maximum, mean, median, and the $10^{\text{th}}$ and $90^{\text{th}}$ percentiles. Sorted by mean time.}
\resizebox{\columnwidth}{!}{%
\begin{tabular}{|l|c|c|c|c|c|c|}
\hline
\textbf{Category} & \textbf{Min} & \textbf{Max} & \textbf{Mean} & \textbf{Median} & \textbf{$P_{10}$} & \textbf{$P_{90}$} \\
\hline
\textbf{Mechanisms} & 0.000035 & 1022.12 & 19.81 & 0.62 & 0.0231 & 17.74 \\
\makecell[l]{Development\\ \& Maintenance} & 0.000012 & 2201.83 & 20.50 & 0.70 & 0.00279 & 35.01 \\
Documentation & 0.0 & 1595.29 & 20.69 & 0.60 & 0.00161 & 29.89 \\
\makecell[l]{User Experience\\ \& Interaction} & 0.000012 & 2537.03 & 27.50 & 0.87 & 0.00375 & 46.84 \\
Infrastructure & 0.0 & 2218.66 & 28.82 & 0.98 & 0.00632 & 52.36 \\
\makecell[l]{Community\\ Interaction} & 0.000046 & 1935.54 & 32.00 & 1.21 & 0.01012 & 71.79 \\
Networking & 0.000035 & 2319.35 & 32.34 & 0.96 & 0.00589 & 67.80 \\
Digital Asset & 0.0 & 2057.94 & 32.55 & 1.14 & 0.00786 & 59.80 \\
Transaction & 0.000035 & 2474.69 & 36.65 & 1.80 & 0.01126 & 76.33 \\
\makecell[l]{Security\\ \& Data Integrity} & 0.000035 & 2446.68 & 45.47 & 2.93 & 0.01337 & 117.87 \\
\textbf{Wallet} & 0.000023 & 2896.19 & 45.95 & 2.02 & 0.00856 & 98.06 \\
\hline
\end{tabular}
}
\label{tab:resolution_times}
\end{table}

Resolution times vary widely across subcategories, from minutes to several years. The longest maximum resolution time occurs in the \textit{Wallet} category (7.9 years), while the shortest is in \textit{Mechanisms} (2.8 years), with most subcategories ranging from 4 to 7 years. Mean resolution times span from 19.8 days (\textit{Mechanisms}) to 46 days (\textit{Wallet}), indicating that issues in categories such as \textit{Wallet}, \textit{Security \& Data Integrity}, and \textit{Transaction} are more resource-intensive or complex. Median resolution times, however, remain lower, ranging from 0.6 days (\textit{Documentation}) to 2.9 days (\textit{Security \& Data Integrity}), showing that 50\% of issues are addressed promptly, while a small share of long-standing cases skew the mean. Such prolonged cases are consistent with prior findings linking delays to vague issue descriptions, limited reproducibility, or shifting project priorities \cite{wan_bug_2017, thung_empirical_2012, lamkanfi_filtering_2012}. The percentile data further illustrates this skew. At the 10th percentile, resolution times are consistently below one hour across all subcategories. By contrast, the 90th percentile ranges from 17.7 days (\textit{Mechanisms}) to 118 days (\textit{Security \& Data Integrity}).


The histogram in Figure~\ref{fig:resolution_distribution} reinforces these findings. It shows that a large majority of issues, particularly in categories like \textit{Mechanisms}, \textit{Documentation}, and \textit{Development \& Maintenance}, are resolved within a month. Over 85\% of issues in most subcategories meet this threshold, with exceptions like \textit{Community Interaction}, \textit{Wallet}, \textit{Security \& Data Integrity}, and \textit{Transaction}. Approximately 4–15\% are resolved between one month and one year, while 1–4\% take over a year.

These findings illustrate that while many issues are resolved quickly, subcategories related to blockchain, especially \textit{Wallets}, require significantly longer attention and resources.

\begin{figure*}[h]
    \centering
    \includegraphics[width=0.85\textwidth]{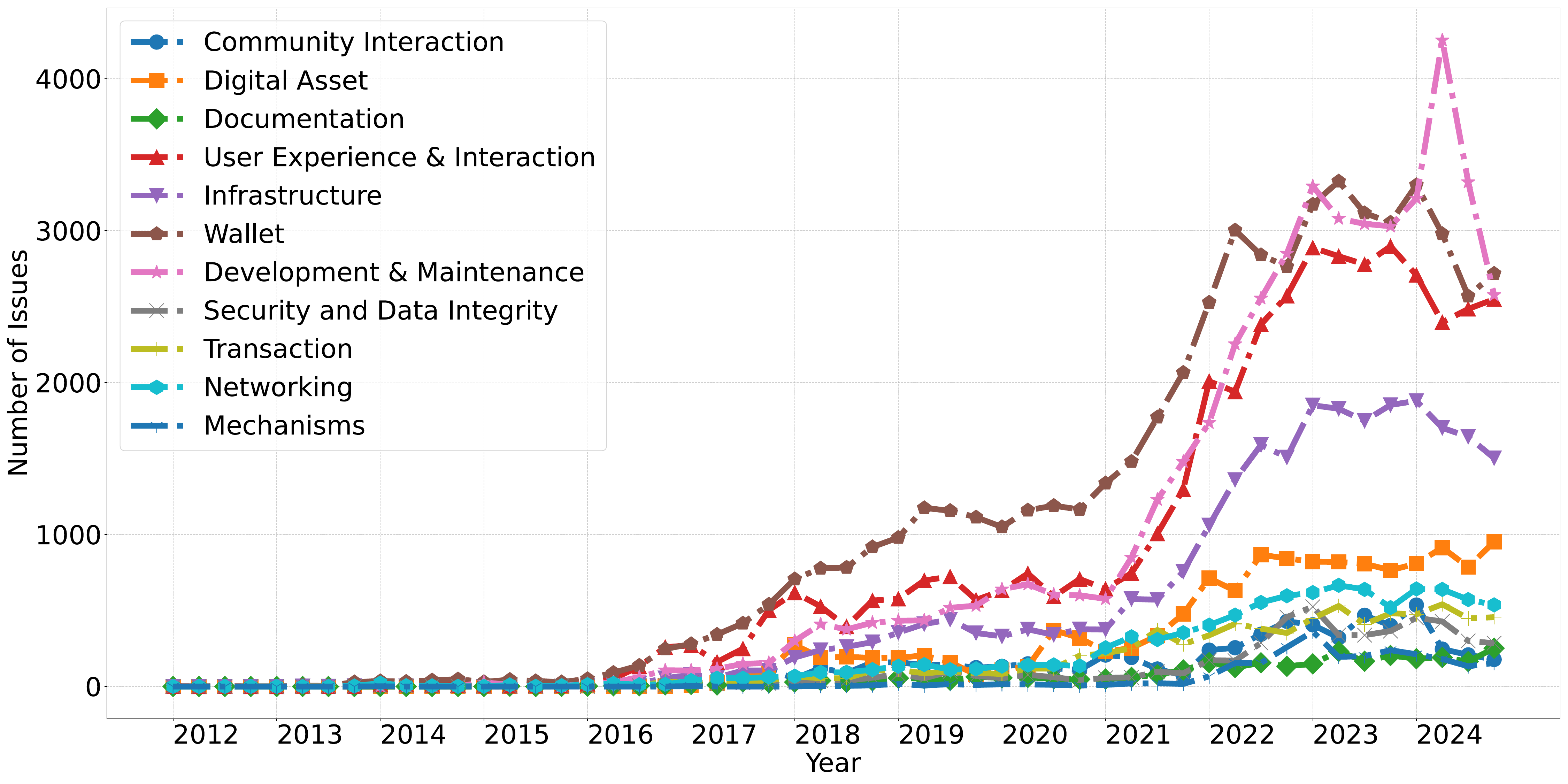}
    \caption{Trends in reported issues by subcategory, quarterly from 2012 to 2024.}
    \label{fig:issue_trends}
\end{figure*}

\begin{summarybox}
Resolution times vary remarkably across issue subcategories: \textit{Wallet} issues take the longest, while \textit{Mechanisms} issues are resolved the fastest. Nonetheless, most issues (around 80\%) across all subcategories are resolved within a month.   
\end{summarybox}
\subsection*{\textbf{RQ3:} How have different types of issues evolved over time in terms of frequency within blockchain projects?}

To understand how issue categories evolved, we analyzed quarterly issue counts for each of the 11 subcategories from 2012 to 2024. Figure~\ref{fig:issue_trends} visualizes these trends, capturing the trajectory of reported issues across the blockchain ecosystem.

Our analysis reveals a statistically significant variation in issue frequencies across subcategories, as confirmed by a Kruskal-Wallis H-test: \textit{\(H=93.67\), \(p = 9.97\times10^{-16}\)}. This non-parametric result indicates that the differences in distributions among the categories are not due to random chance, underscoring the heterogeneity in issue evolution.


A significant rise in issue frequency is seen across nearly all subcategories beginning in 2016. A similar trend has also been observed in previous studies on dApps~\cite{wu_first_2021, leiponen_dapp_2022}. This sharp increase coincides with the launch and adoption of platforms like Ethereum, which officially released its Homestead version in March 2016. Ethereum introduced a new paradigm of decentralized applications (dApps) and smart contracts, significantly expanding the scope of blockchain software development. This growth spurred the creation of many blockchain projects, which in turn generated an increasing number of reported issues, particularly in categories like \textit{Development \& Maintenance}, \textit{User Experience \& Interaction}, and \textit{Wallet}.

From 2023 onward, we observed a noticeable decline in issue frequency across most subcategories. This trend may stem from ecosystem maturation, better documentation, reusable tools that reduce the need for extensive issue tracking, or a broader shift of open-source efforts toward AI-related projects, though further research is needed. Despite the decline, subcategories like \textit{Wallet} and \textit{UI} continue to exhibit high activity, reflecting their role in user interaction and asset management.

Among the individual categories, \textit{Development \& Maintenance} and \textit{Wallet} show the strongest upward trends, suggesting these areas have received increasing focus. \textit{User Experience \& Interaction} and \textit{Infrastructure} also exhibit strong growth, although they plateau slightly below the top-performing subcategories. In contrast, \textit{Documentation} and \textit{Mechanisms} lag behind, maintaining relatively lower positions throughout the timeline, indicating either slower progress or less emphasis. \textit{Community Interaction}, \textit{Digital Asset}, and \textit{Networking} show steady progress, while \textit{Security \& Data Integrity} and \textit{Transaction} fall in the mid-range with slightly fluctuating but overall upward trends. Overall, the trends show broad and varied characteristics of different subcategories, with some leading the development trajectory more than others.

\begin{summarybox}
The frequency of issues surged after 2016, coinciding with the rise of Ethereum and dApps, then started declining after 2022. \textit{Wallet}, the most frequently reported subcategory, consistently remained the most reported throughout the timeline.
\end{summarybox}
\section{Discussion}
\label{sec:discussion}

This study provides a large-scale empirical analysis of open-source blockchain software projects using GitHub issue mining and BERTopic modeling. Findings show that \textit{Development \& Maintenance} and \textit{Wallet}-related issues dominate, with nearly half being blockchain-specific. Resolution time analysis reveals that \textit{Security} and \textit{Wallet} issues take the longest to resolve, making them particularly resource-demanding. Trend analysis further indicates that issue frequency grew during the rise of Ethereum and dApps after 2016 but started declining after 2022.

The findings have important implications for both research and practice. First, the predominance of blockchain-specific issues (nearly 50\%) demonstrates that conventional software engineering practices alone cannot adequately address the unique challenges of decentralized systems. Practitioners must adopt blockchain-tailored approaches to issue tracking and triaging. In particular, the longer resolution times in max and 90th percentile for \textit{Security} and \textit{Wallet} issues highlight the need for dedicated resources, stronger expertise, and specialized frameworks. By prioritizing based on complexity, project maintainers can improve efficiency and reduce bottlenecks.

Second, the dominance of \textit{Wallet} issues underscores their technical and security complexity. Wallets lie at the intersection of usability, interoperability, and security, making them vulnerable to long-standing challenges. This suggests a pressing need for standardized frameworks, automated testing strategies, and reusable tools to streamline wallet development.

Third, the temporal analysis illustrates how platform innovations shape issue spaces. The surge of issues after 2016 and the decline after 2022 suggest that issue patterns reflect ecosystem shifts, likely influenced by the introduction of Ethereum and decentralized applications, and later by a global focus shift toward artificial intelligence. Researchers should therefore pursue longitudinal studies to examine how practices, tools, and issue categories evolve in response to such shifts.

Finally, the emergence of novel issue types, such as governance mechanisms and community interaction, indicates that blockchain development will continue diversifying rather than stabilizing. This dynamic nature makes continuous developer training essential, ensuring practitioners are prepared for new application domains and problem spaces.

Overall, this study shows that blockchain projects need methodologies, tooling, and expertise beyond traditional practices. For practitioners, the findings suggest prioritizing resources on \textit{Wallet Management}, \textit{Security}, and \textit{Infrastructure}. For researchers, they offer a foundation for developing specialized tools and longitudinal studies. Addressing these needs can improve the security, maintainability, and usability of blockchain systems as they expand into new domains.





\subsection{Limitations \& Future Work}
This study has several limitations that suggest directions for future research. First, its focus on open-source projects may not fully capture the breadth of development practices and challenges across the broader blockchain ecosystem. Another limitation arises from relying solely on issue titles for topic modeling, as they often contain auto-generated or template text in issue bodies. However, many issues include rich, developer-written content that could provide deeper insights. 

Future research could address these limitations by incorporating closed-source blockchain projects to compare practices across different development environments. Expanding the scope to include non-blockchain projects would help highlight challenges unique to blockchain technology. Exploring distinct blockchain domains, such as cryptocurrency, healthcare, finance, supply chain, and smart contracts, would provide a more comprehensive view of industry-specific issues. Finally, the quality of generated topics could be improved by incorporating informative issue bodies while using automated filtering to exclude templated or auto-generated content.
\section{Conclusion}
\label{sec:conclusion}
This study investigates the challenges in blockchain-based software applications by mining and analyzing roughly half a million GitHub issues from more than a thousand open-source projects. Using topic modeling with BERTopic followed by manual annotation, 49 distinct topics were identified, which were further organized into a hierarchical structure comprising 11 major subcategories. General software issues slightly outnumber blockchain-specific ones, with frequent concerns including logging, wallet management, and UI enhancement. Issue resolution times varied, with wallet-related issues taking longer. While issue activity peaked around 2016 and started declining after 2022, wallet issues remained consistently significant. The results show that blockchain development shares challenges with general software engineering but also faces unique issues that require dedicated solutions and practices.

\bibliographystyle{IEEEtran}  
\bibliography{references} 
\end{document}